\documentclass[floats,preprint,aps,tightenlines,showpacs,nofootinbib]{revtex4}

\usepackage{latexsym}
\usepackage[dvips]{graphics}

\newcommand{\del}{\partial}

\newcommand{\beq}{\begin{equation}}
\newcommand{\eeq}{\end{equation}}
\newcommand{\beqa}{\begin{eqnarray}}
\newcommand{\eeqa}{\end{eqnarray}}
\newcommand{\sr}{\sqrt}
\newcommand{\fr}{\frac}

\newcommand{\vp}{\varphi}

\begin{document}

\preprint{ hep-th/0401225, KIAS-P03062 }

\title{Classical Stability of Black D3-branes}

\author{Gungwon Kang$^{a,}$\footnote{E-mail address: gwkang@kias.re.kr} and
Jungjai Lee $^{b,c,}$\footnote{E-mail address:
jjlee@daejin.ac.kr}}

\affiliation{$^a$School of Physics, Korea Institute for Advanced
Study, 207-43 Cheongryangri-dong, Dongdaemun-gu, Seoul 130-012, Korea \\
$^b$Department of Physics, Daejin University, Pocheon, Gyonggi
487-711, Korea \\
$^c$Department of Physics, North Carolina State University,
Raleigh, North Carolina 27695-8202, USA}

\begin{abstract}

We have investigated the classical stability of charged black
$D3$-branes in type IIB supergravity under small perturbations.
For $s$-wave perturbations it turns out that black $D3$-branes are
unstable when they have small charge density. As the charge
density increases for given mass density, however, the instability
decreases down to zero at a certain finite value of the charge
density, and then black $D3$-branes become stable all the way down
to the extremal point. It has also been shown that such critical
value at which its stability behavior changes agrees very well
with the predicted one by the thermodynamic stability behavior of
the corresponding black hole system through the Gubser-Mitra
conjecture. Unstable mode solutions we found involve non-vanishing
fluctuations of the self-dual five-form field strength. Some
implications of our results are also discussed.

\end{abstract}


\maketitle

\newpage

\section{Introduction}

It is well known that the four-dimensional Schwarzschild black
hole in Einstein gravity is stable classically under linearized
perturbations. Recently, Ishibashi and
Kodama~\cite{Ishibashi:2003ap} have shown that this stable
behavior extends to hold for higher dimensional cases. However,
some black strings or branes, which have hypercylindrical horizons
instead of compact hyperspherical ones, are found to be unstable
as the compactification scale of extended directions becomes
larger than the order of the horizon radius - the so-called
Gregory-Laflamme instability~\cite{GL1}. The simplest black string
would be the five-dimensional Schwarzschild black string in
Einstein gravity that is a foliation of four-dimensional
Schwarzschild black holes along the fifth direction.

Gregory and Laflamme~\cite{Gregory:1994bj} also considered a class
of magnetically charged black $p$-brane solutions for a stringy
action containing the NS5-brane of the type II supergravity. For
horizons with infinite extent, they have shown that the
instability persists to appear but decreases as the charge
increases to the extremal value. On the other hand, branes with
extremal charge turned out to be stable~\cite{Gregory:1994tw}.
Since their discovery of such linearized instability, black
strings or branes have been believed to be generically unstable
classically under small perturbations except for the cases of
extremal or suitably compactified ones, and the Gregory-Laflamme
instability has been used to understand physical behaviors of
various systems involving black brane configurations as in string
theory. Recently, however, Hirayama and
Kang~\cite{Hirayama:2001bi} analyzed the stability of three types
of black string backgrounds in five-dimensional AdS space. With or
without the presence of a 3-brane, the geometry of these black
strings in consideration is warped in the fifth direction,
resulting in no translational symmetry along the horizon. They
showed that the AdS$_4$-Schwarzschild black string becomes stable
as the horizon radius is larger than the order of the AdS$_4$
radius whereas Schwarzschild~\cite{Gregory:2000gf} and
dS$_4$-Schwarzschild black strings are unstable as usual. It is
possible to have stationary black string or brane solutions even
in four dimensions when a negative cosmological constant is
present. Interestingly it seems that all known stationary black
branes in four dimensions are stable. In particular, the case of
BTZ black strings has been checked explicitly~\cite{Kang:2002hx}.

In the context of string theory black branes that Gregory and
Laflamme considered are those having magnetic charges with respect
to Neveu-Schwarz gauge fields only~\cite{Gregory:1994bj}. Having
found some black string systems in which the Gregory-Laflamme
instability is absent as mentioned above, Hirayama, Kang, and
Lee~\cite{Hirayama:2002hn} have also considered a wider class of
black brane solutions for string gravity in order to see whether
or not the stability behavior drastically changes. Indeed it turns
out that the stability of black branes behaves very differently
depending on the parameter $a$ that specifies the strength of
coupling between dilaton and gauge fields in the theory as in
Eq.~(\ref{action}) below. That is, for magnetically charged static
black brane solutions in theories of this form~\cite{Duff:1996hp},
there exists a critical value of the coupling parameter $a_{\rm
cr}(D,p)$ to be determined by the full spacetime dimension $D$ and
the dimension of the spatial worldvolume $p$ of those black
$p$-branes. The case that Gregory and Laflamme studied is
precisely when $a=a_{\rm cr}$. Black branes with horizons of
infinite extent in this case are always unstable as explained
above, and magnetically charged NS5-branes of the type II
supergravity belong to this class. When $0 < a < a_{\rm cr}$,
black branes with small charge are unstable as usual. As the
charge increases, however, the instability decreases and
eventually disappears at a certain critical value of the charge
density which could be even far from the extremal point.
Magnetically charged black D0, F1, D1, D2, D4 branes of the type
II string theory belong to this class for instance. When $a >
a_{\rm cr}$, on the other hand, the instability persists all the
way down to the extremal point. Magnetically charged black D5 and
D6 branes are in this case for example. However it is shown that
all black branes mentioned above are stable at the extremal point,
which might be expected due to the BPS nature of extremal
solutions in string theory.

Recently, Gubser and Mitra~\cite{Gubser:2000ec} gave a conjecture
about when the classical instability of a black brane sets on in
terms of thermodynamic stability. This Gubser-Mitra (GM)
conjecture is a sort of refinement of the entropy comparison
argument given by Gregory and Laflamme~\cite{GL1,Gregory:1994bj},
and states that a black brane with a non-compact translational
symmetry is classically stable if and only if it is locally
thermodynamically stable. As argued by Reall~\cite{Reall:2001ag},
when we expand classical perturbations in terms of Fourier modes
along the horizon having translational symmetry, the set of
linearized perturbation equations for a black brane becomes the
Lichnerowitz equation with an additional mass term for the black
hole on horizon cross sections. Now it can be seen that the
existence of a threshold mode for instability of a black brane is
related to the presence of a negative eigen mode in the Euclidean
path integral for the black hole system. Consequently, the
partition function gets an imaginary contribution, implying a
thermodynamic instability of the black hole system on the horizon
cross sections, and vice versa. This interesting relationship
between classical dynamical and local thermodynamic stabilities
has been checked explicitly for various black string or brane
systems~\cite{Gubser:2000ec,Kang:2002hx,Hirayama:2002hn,PGR,Gubser:2002yi}.
When the translational symmetry along the horizon is broken, one
can see some disagreements for on set points for instability as
shown in the stability analysis for AdS$_4$-Schwarzschild black
strings in AdS space~\cite{Hirayama:2001bi}. It also should be
pointed out that this conjecture simply gives the information
about when a black string or brane becomes stable or unstable. It
does not explain or predict other details of classical stability
behaviors~\cite{Hirayama:2002hn}.

In the present paper, we analyze the classical stability of
charged static black brane solutions for the theory in
Eq.~(\ref{action}) in the case that $n=D/2$ ({\it i.e.}, $a=0$)
with a self-dual $n$-form field strength. This case includes black
D3-branes in the type II supergravity, and is of interest for
several reasons. Firstly, note that the geometry of the spatial
worldvolume of black brane backgrounds whose linearized
stabilities have been analyzed so far in the literature is flat.
As can be seen in Eq.~(\ref{pbrane}) below, however, black brane
backgrounds to be considered in this paper do not have flat
spatial worldvolume, but have a warping factor multiplied. Such
overall factor can not be removed by finding a suitable
conformally equivalent theory as in
Refs.~\cite{Reall:2001ag,Hirayama:2002hn} since the background
dilaton field is constant in this case. Therefore a sort of
Kaluza-Klein reduction of perturbation equations for these black
{\it branes} does not give the standard form of the Lichnerowitz
equation with an additional mass term for black {\it holes} on the
horizon cross sections as usual. Secondly, in the $s$-wave
perturbation analyses in
Refs.~\cite{Reall:2001ag,Hirayama:2002hn}, fluctuations of the
field strength for unstable modes could be set to be zero
consistently partly because black branes are charged magnetically
only. However, black brane backgrounds in the consideration are
charged electrically as well as magnetically since the five-form
field strength is self-dual. Subsequently, it is not consistent to
set $s$-wave fluctuations of the field strength being frozen as
shall be shown below explicitly. Finally, the case of black
$D3$-brane is not included in the proof of the GM conjecture by
Reall~\cite{Reall:2001ag}. Although he suggests some
generalization of the argument in Ref.~\cite{Reall:2001ag} could
include the case, such generalization seems to be non-trivial for
the reasons mentioned above. Moreover, the covariant action
including a self-dual field strength is not known, and so it is
not clear how the self-duality condition could be incorporated in
the proof. Therefore it is interesting to see not only how
different features mentioned above affect the detailed stability
analysis for black branes to be considered in this paper, but also
whether or not the GM conjecture still holds for these ``dyonic''
branes.

Actually the classical stability of the black D3-brane has been
studied by using a notion of universality classes recently by
Gubser and Ozakin~\cite{Gubser:2002yi}. For s-wave fluctuations
they dimensionally reduce the ten-dimensional type IIB
supergravity action to a three-dimensional one that contains
gravity and two scalar fields only, and perform there the
stability analysis for static perturbations in a certain specific
form. However it is not clear whether or not all relevant
perturbations in the original theory have been covered in such
analysis. In particular, the field strength is assumed to be
frozen in their dimensional reduction. Although such approximation
would be of little significance for small charge, it might change
the stability behavior significantly as the charge increases.
Finally, we would like to point out that in string theory the
system of D3-branes is understood best in the context of the
AdS/CFT correspondence. Hence the details of stability behavior in
gravity side might be very useful for understanding corresponding
behaviors in the CFT side.

In section \ref{thermo}, we briefly summarize the local
thermodynamic behaviors of black D3-branes in order to get a hint
at the classical stability predicted by the Gubser-Mitra
conjecture. In section \ref{PA}, we perform the perturbation
analysis explicitly and give numerical results. Finally, some
possible physical implications of our results are discussed.

\section{Thermodynamic behavior}
\label{thermo}

Let us consider the action given by
 \beqa
{\rm I} &=& \int d^Dx\sr{-\bar{g}} \left[ e^{-2 \bar{\phi}} \left(
\bar{R} +4 (\partial \bar{\phi})^2 \right) -\fr{1}{2n!}F_n^2
\right]
\nonumber \\
&=& \int d^Dx\sr{-g} \left[ R -\fr{1}{2} (\partial \phi)^2
-\fr{1}{2n!}e^{a\phi}F_n^2 \right].
 \label{action}
 \eeqa
The second form is written in Einstein frame with $a=
(D-2n)/\sr{2(D-2)}$. It is known that there is no covariant action
for low energy type IIB supergravity due to the self-duality
condition for the field strength ${\bf F}_5 = \mbox{}^{*}{\bf
F}_5$. However, the action in Eq.~(\ref{action}) with $D=10$ and
$n=5$ is quite close to the type IIB supergravity with all other
form fields set to be zero. When the rank of the field strength is
$n=D/2$, one has $a=0$ and the field equations for the action in
Eq.~(\ref{action}) are given by
 \beqa
&& \nabla_M F^{MP_1 \cdots P_{n-1}} =0, \qquad\qquad
 \nabla^2 \phi =0,  \nonumber \\
&& R_{MN} = \frac{1}{2} \nabla_M \phi \nabla_N \phi +
\frac{1}{2(n-1)!} F_{MP_1 \cdots P_{n-1} } F_N^{P_1 \cdots
P_{n-1}} - \frac{1}{4n !} F^2 g_{MN} .
 \label{eom}
 \eeqa
In the following we consider a gravity theory in which the
dynamics of fields is governed by the field equations given in
Eq.~(\ref{eom}) with an additional constraint of self-duality for
the field strength on solutions such as
 \beq
 {\bf F} = \mbox{}^*{\bf F} .
 \label{self-duality}
 \eeq

Static black $p$-brane solutions for this theory
are~\cite{Duff:1996hp}
 \beqa
&& ds^2 = -U dt^2 +\fr{dr^2}{U} +R^2 d\Omega_n^2 +W
\delta_{ij}dz^i dz^j  \nonumber \\
&& \qquad  = W \left( -N dt^2 +\delta_{ij}dz^i dz^j
\right) + W^{-1} \left( N^{-1} dr^2 +r^2 d\Omega_n^2 \right),  \nonumber  \\
&& \phi = 0, \qquad\qquad  {\bf F} = \lambda
\left(\mbox{\boldmath$\epsilon$}_n +
\mbox{}^*\mbox{\boldmath$\epsilon$}_n \right),
 \label{pbrane}
 \eeqa
where
 \beq
N= 1-\fr{k}{r^{n-1}} , \quad W=\left( 1+\fr{k}{r^{n-1}} \sinh^2
\mu \right)^{-2/(n-1)},
 \quad
U = WN  ,
 \quad
R^2 = r^2 W^{-1} .
 \eeq
Here $n=D/2=p+2$, the $p$-dimensional spatial worldvolume
directions are denoted by $z^i$ with $i=1,2,\cdots, p$, and
$\mbox{\boldmath$\epsilon$}_n$ is the volume form on the unit
sphere $d\Omega_n^2$. The mass and both electric and magnetic
charge densities are
 \beq
M = k\left(n +4 \sinh^2 \mu \right), \qquad \lambda =
\sr{\fr{n-1}{2}} k \sinh 2\mu ,
 \label{ADM}
 \eeq
respectively.\footnote{Note here that the value of the charge
density is corrected from the one given in Ref.~\cite{Duff:1996hp}
({\it i.e.}, $\lambda = k\sinh 2\mu /\sr{2}$).} Notice that the
spatial worldvolume of this brane is not flat except for the
uncharged case ({\it i.e.}, $W=1$). The maximally charged extremal
limit is $k \rightarrow 0$ and $\mu \rightarrow \infty$ with the
mass density (or, $ke^{2\mu}$) fixed.

Now let us consider a finite segment of this black $p$-brane with
unit worldvolume. Being regarded as a thermal system, it has
entropy and temperature given by
 \beq
 S \sim \left( \cosh \mu \right)^{\fr{4}{n-1}}
r_{\scriptscriptstyle{H}}^{n}  \quad {\rm and} \quad T=\fr{n-1}{4
\pi r_{\scriptscriptstyle{H}}} \left( \cosh\mu
\right)^{-\fr{4}{n-1}} ,
 \label{ST}
 \eeq
respectively. Here $r_{\scriptscriptstyle{H}} = k^{1/(n-1)}$ is
the horizon radius. The specific heat capacity is given by
 \beq
 C_{\lambda} =\left( \fr{\del M}{\del T}\right)_{\lambda} = -4 \pi
r_{\scriptscriptstyle{H}}^{n} \left( \cosh \mu
\right)^{\fr{4}{n-1}}\,\, \fr{2+(n-2)\cosh 2\mu}{1 -2\sinh^2 \mu}.
 \label{heatcapacity}
 \eeq
Here one finds that, as $\mu$ increases, the heat capacity changes
its sign from negativeness to positiveness at a certain critical
value of $\mu$ given by
 \beq
 \sinh^2 \mu_{\rm cr} = 1/2 .
 \label{criticalmu}
 \eeq
Thus, if the Gubser-Mitra conjecture holds for this system as
well, we expect that these black $p$-brane backgrounds become
stable classically under small perturbations for $\mu \ge \mu_{\rm
cr}$. Defining an extremality parameter as
 \beq
 q = \fr{\lambda}{\lambda_{\rm max}} = \fr{2 \sinh 2\mu}{n+4\sinh^2 \mu}
 \label{nonextrem}
 \eeq
with $\lambda_{\rm max} = \sr{(n-1)/8}\, M$, one notice that this
critical value corresponds to
 \beq
 q_{\rm cr} = \fr{2\sr{3}}{n+2} .
 \label{criticalq}
 \eeq
For the black D3-brane ({\it i.e.}, $n=5$) this value is $q_{\rm
cr} = 2\sr{3}/7 \simeq 0.495$.

\section{Perturbation analysis}
\label{PA}

In this section we perform the classical stability analysis for
small perturbations of fields at the linear level. Under $\phi
\rightarrow \phi +\delta \phi$, ${\mathbf F} \rightarrow {\mathbf
F} +\delta {\mathbf F}$, and $g_{MN} \rightarrow g_{MN} +h_{MN}$,
we have from Eq.~(\ref{eom}) linearized perturbation equations
given by
 \beqa
 \mbox{} && \nabla^2 \delta \phi - \nabla_{(M} \nabla_{N)} \phi
h^{MN} - \nabla_{(M} \phi \nabla_{N)} \Big( h^{MN} - \frac{1}{2}
g^{MN} h \Big) =0 ,  \label{leomD}  \\
&& \nabla_M \delta F^{MP_1 \cdots P_{n-1} }-\nabla_M
F^{QP_1
\cdots P_{n-1}} h^M_Q \nonumber \\
&& -(n-1) F^{MQ[P_2 \cdot P_{n-1}} \nabla_M h^{P_1]}_Q
 -\nabla_M \Big( h^M_Q - \frac{1}{2} h \delta^M_Q \Big) F^{Q P_1 \cdot
P_{n-1}} =0 ,  \label{leomF} \\
&& \nabla^2 h^M_N + \nabla^M \nabla_N h - \nabla_Q \nabla^M h^Q_N
- \nabla^Q \nabla_N h^M_Q + 2 \nabla^{(M} \phi \nabla_{N)} \delta
\phi  \nonumber \\
&& +\frac{1}{(n-1)!} \bigg( 2\delta F^{(M |P_1 \cdots P_{n-1}|}
F_{N)P_1 \cdots P_{n-1}}-(n-1) F^M_{PR_1 \cdot R_{n-2} } F^{QR_1
\cdots R_{n-2}} h^P_Q \bigg) \nonumber \\
&& -\frac{1}{2n!} \bigg[ F^2 h^M_N + \delta^M_N \Big( 2F^{p_1
\cdots P_n} \delta F_{P_1 \cdots P_n} - n F_{PR_1 \cdots R_{n-1}}
F^{QR_1
\cdots R_{n-1}} h^P_Q \Big) \bigg] =0
 \label{leomG}
 \end{eqnarray}
with the perturbed self-duality condition
\begin{eqnarray}
\delta ({\mathbf F} - \mbox{}^*\!{\mathbf F} )=0 .
\label{self-dualityL}
\end{eqnarray}
It is very important to notice that the self-duality condition
imposed on the perturbed field strength in
Eq.~(\ref{self-dualityL}) is not $\delta {\mathbf F} =
\mbox{}^*\!\delta {\mathbf F}$, but
 \begin{eqnarray}
\delta F_{M_1 \cdots M_n}= (\mbox{}^*\!\delta F )_{M_1 \cdots M_n
} + \frac{1}{2} h F_{M_1 \cdots M_n} - \frac{1}{(n-1)!} h^{PQ}
F^{P_1 \cdots P_{n-1}}_Q \epsilon_{PP_1 \cdots P_{n-1} M_1 \cdots
M_n} .
 \label{self-dualityL2}
\end{eqnarray}
In addition the Bianchi identity requires
 \beq
 d\,\delta {\mathbf F} =0.
 \label{Bianchi}
 \eeq

In order to see whether or not the black $p$-brane backgrounds in
Eq.~(\ref{pbrane}) are stable under small perturbations, we need
to check if there is any solution for linearized equations given
above that is growing in time while it is regular spatially
outside the event horizon. In case that the black brane is stable,
it would be very difficult to show there exists no such unstable
solution in general. When the black $p$-brane background is
unstable, however, it is rather easy to find a certain type of
unstable solutions. Notice first that, since the background
dilaton field is constant, the fluctuation of the dilaton field
can be seen from Eq.~(\ref{leomD}) to be completely decoupled,
{\it i.e.}, $\nabla^2 \delta \phi = 0$. Hence one can set $\delta
\phi =0$. Since the background fields in Eq.~(\ref{pbrane}) are
stationary and invariant under translations in directions of the
spatial worldvolume, one can also assume that
 \beq
\delta {\mathbf F}(x^{\mu},z^i) = e^{\Omega t +im_iz^i} \delta
{\mathbf F}(r,\theta^m) , \qquad h_{MN}(x^{\mu},z^i) = e^{\Omega t
+im_iz^i} H_{MN}(r,\theta^m)
 \label{ansatz1}
 \eeq
for unstable mode solutions. Here $\Omega > 0$ and $\{ \theta^m
\}$ are angular coordinates for the $n$-sphere. We denote the
coordinate system by $\{ x^M \} =\{ x^{\mu},z^i \} =\{
t,r,\theta^m,z^i \}$.

For $s$-wave perturbations that are spherically symmetric in
submanifolds perpendicular to the $p$-dimensional spatial
worldvolume, one can easily see that the Bianchi identity in
Eq.~(\ref{Bianchi}) restricts the form of $\delta {\mathbf F}$
further, yielding the only non-vanishing components given by
 \begin{eqnarray}
\delta F_{trz^1 \cdots z^p} = e^{ \Omega t + i m_i z^i} Q(r).
 \label{PF}
 \end{eqnarray}
In order to see the presence of unstable mode solutions, it
suffices to consider the threshold modes ({\it i.e.}, $\Omega =0$)
only as usual~\cite{PGR,Hirayama:2002hn,Reall:2001ag}. Note that
we have not imposed any gauge such as the transverse gauge ({\it
e.g.}, $\nabla_M h^{MN} = \nabla^N h/2$) for metric perturbations
$h_{MN}$ in Eq.~(\ref{leomG}) so far. For unstable threshold modes
one may set $H_{tr}=H_{ti}=0$ by choosing a suitable gauge as in
Refs.~\cite{PGR,Reall:2001ag}, instead of the usual transverse
gauge as in Gregory and Laflamme~\cite{Gregory:1994bj}. Now the
remaining non-vanishing components of $H_{MN}$ are $H_{tt}$,
$H_{rr}$, $H_{ri}$, $H_{\theta \theta}$ and $H_{ij}$. By looking
at the form of perturbation equations for $H_{ij}$ and $H_{ri}$,
one may have
\begin{eqnarray}
H^i_j = \frac{m_i m_j}{m^2} \zeta (r) + \delta^i_j \rho (r)
\end{eqnarray}
and
\begin{eqnarray}
H^r_i = i \frac{m_i}{m} \eta (r) ,
\end{eqnarray}
where $m^2= \sum^p_{i=1}m^2_i$. If $\rho =0$, this is exactly the
same ansatz used by Gregory and Laflamme~\cite{Gregory:1994bj}.
The presence of the new term $\rho$ above can be understood as
follows. Under the diffeomorphisms associated with a vector field
$\xi^M=e^{i m_i z^i} \xi^M(r)$, we have
\begin{eqnarray}
\tilde{H}_{ij} = H_{ij} -im_i\xi_j(r) -im_j \xi_i(r) -\delta_{ij}
V\left( \frac{W'}{W} \right)\xi_r(r) .
\end{eqnarray}
Since $W'=\partial_r W$ does not vanish for charged black branes,
the term $\rho$ should appear in general. On the other hand, the
$H_{ri}$ component transforms as
 \beq
 \tilde{H}_{ri} = H_{ri} -im_i\xi_r(r) -\xi'_i(r)
 +\fr{W'}{W} \xi_i(r) .
 \eeq
Therefore, when the spatial worldvolume of a black brane is a
warped flat space as in the present case, one sees that only two
of three functions $\zeta$, $\rho$ and $\eta$ can be eliminated by
choosing suitable functions $\xi_r(r)$ and $\xi(r)$ in $\xi_i(r) =
im_i \xi(r)$. We set $\zeta = \eta =0$ in our analysis.
Consequently, for threshold modes the non-vanishing components of
$s$-wave metric perturbations finally  become
\begin{eqnarray}
h^M_N &=& e^{i m_i z^i} {\rm diag} \left(H^t_t , H^r_r ,
H^{\theta^m}_{\theta^m}, H^i_i \right)  \nonumber \\
&=& e^{i m_i z^i} {\rm diag} \left(\varphi, \psi , \chi, \cdots,
\chi, \rho, \cdots, \rho \right) ,
 \label{PH}
 \end{eqnarray}
where all unknown functions $\varphi, \psi , \chi,$ and $\rho$ are
functions of the radial coordinate $r$ only.

In Appendix A, all linearized perturbation equations in
Eqs.~(\ref{leomF}) and (\ref{leomG}) are shown in components. We
have seven coupled ordinary differential equations for seven
unknown functions $\varphi$, $\psi$, $\chi$, $\zeta$, $\eta$,
$\rho$ and $Q$. Although we have not used the transverse gauge
condition, note that in the Reall gauge ({\it i.e.}, $\zeta =
\eta=0$) the linearized perturbation equation Eq.~(\ref{leomiNj})
becomes equivalent to the $z^i$-component of the transverse gauge
Eq.~(\ref{TGi}). As can be seen in Eq.~(\ref{leomTG}), however,
Eq.~(\ref{leomri}) differs from the $r$-component of the
transverse gauge Eq.~(\ref{TGr}) as the black brane gets charged
({\it i.e.}, $W' \not= 0$). This property is different from the
cases studied in Refs.~\cite{Hirayama:2002hn,Reall:2001ag} where
some of linearized perturbation equations in the Reall gauge
become precisely equal to the transverse gauge.

On the other hand, the only non-trivial component of the
self-duality constraint in Eq.(\ref{self-dualityL2}) is
\begin{eqnarray}
2Q - \lambda \left( \varphi + \psi +(n-2)\rho-n\chi +\zeta \right)
=0 .
 \label{self-dualityL3}
\end{eqnarray}
It is very interesting to see that this equation is equivalent to
the linearized equations for gauge fields in Eq.~(\ref{leomF}) as
can be seen in Eqs.~(\ref{leomFcomp1}) and (\ref{leomFcomp2}).
Thus the self-duality constraint is automatically satisfied at
least for any $s$-wave threshold mode solutions of the linearized
equations. Eq.~(\ref{self-dualityL3}) shows that the fluctuation
of the field strength is proportional to the charge density when
the brane is charged weakly. It also implies that $\delta {\mathbf
F} \sim Q$ cannot be frozen to be zero for a finitely charged
brane since the term in parentheses in Eq.~(\ref{self-dualityL3})
would not vanish in general. For the case of the $D3$-brane with
the gauge $\zeta =\eta =0$ for example, it is enough to see that
the zeroth-order equation in $\lambda$ ({\it i.e.}, $\varphi_0 +
\psi_0 -5\chi_0 =0$) is inconsistent with other uncharged
linearized equations.

In our gauge chosen as above we now have a set of seven equations
for five unknown functions $\varphi$, $\psi$, $\chi$, $\rho$ and
$Q$, which consists of two zeroth-order, one first-order, and four
second-order coupled ordinary differential equations. By doing
lengthy calculations one can show that two of them are not
independent of others, resulting in five independent equations for
five unknowns. From now on we restrict our consideration into the
case of the black $D3$-brane, {\it i.e.}, $n=D/2=p+2=5$. We expect
that the analysis for other cases of black branes in
Eq.~(\ref{pbrane}) can be done similarly and that their stability
behaviors are essentially same as this case. By eliminating the
functions $\chi$, $\rho$ and $Q$, we end up with two second-order
semi-decoupled differential equations for $\varphi$ and $\psi$ as
follows:
 \beqa
 \mbox{} && \Big(r^4 -1\Big)\, \vp''
 +\fr{1}{r K}\Bigg[ m^2r^2\Big(r^4+b\Big)^4\Big(5r^4-1\Big)
 -35r^{20} +(37b+46)r^{16} +\Big(63b^2  \nonumber \\
 && -62b -19\Big)r^{12} +3b\Big(5b^2 -18b +11\Big)r^8
 +b^2(6b+31)r^4 +3b^3
  \Bigg]\, \vp'
-\fr{1}{(r^4-1)K} \Bigg[m^4 \nonumber \\
&& r^2 \Big(r^4-1\Big)\Big(r^4+b\Big)^5 -m^2\Big(r^4+b\Big)^2\Big[
7r^{16} -2(6b+5)r^{12}
-\Big(11b^2 -12b +5\Big)r^8  \nonumber \\
&& +2b(3b-8)r^4 -3b^2 \Big] +16r^2\Big[ (b+2)r^{12} -3b(3b+2)r^8
-3b\Big(2b^2-1\Big)r^4 -b^2 \Big]
\Bigg]\, \vp \nonumber \\
&& +\fr{2}{r K}\Big[7r^8 +(b-5)r^4 +b\Big] \Big[
 (b+2)r^8 -5b(b+1)r^4 -b^2 \Big]\, \psi' +\fr{2}{(r^4-1)K}
\Bigg[ m^2 \nonumber \\
&& \Big(r^4+b\Big)^2\Big[7(b+2)r^{12} -\Big(11b^2 +8b +10\Big)r^8
+b(14b+9)r^4 +b^2\Big] -8r^2 \Big[\Big(10b^2
\nonumber \\
&& +5b -2\Big)r^{12} +b\Big(2b^2 -13b -6\Big)r^8 +b\Big(2b^2 +14b
+3\Big)r^4 +b^2(2b-1)\Big] \Bigg] \, \psi = 0
 \label{SDEvphi}
 \eeqa
and
 \beqa
 \mbox{} && \Big(r^4 -1\Big)\, \psi''
+\fr{1}{r K} \Bigg[ m^2r^2\Big(r^4+b\Big)^4\Big(5r^4-1\Big)
-49r^{20} +(187b+94)r^{16} +\Big(13b^2  \nonumber \\
&& -358b -37\Big)r^{12} -b\Big(7b^2+38b -163\Big)r^8
-3b^2(6b+5)r^4 +b^3 \Bigg] \,
\psi' -\fr{1}{(r^4-1)K}\Bigg[ m^4 \nonumber \\
&& r^2\Big(r^4-1\Big)\Big(r^4+b\Big)^5 +m^2\Big(r^4+b\Big)^2\Big[
7r^{16} -2(30b+11)r^{12} +\Big(21b^2
+140b +23\Big)r^8  \nonumber \\
&& -2b(7b+32)r^4 +b^2 \Big] -16r^2\Big[ 6(4b+1)r^{16} -\Big(2b^2
+63b +4\Big)r^{12} -b\Big(2b^2 +13b -52\Big)r^8 \nonumber \\
&& -b\Big(4b^2 -8b +15\Big)r^4 -3b^2 \Big] \Bigg] \, \psi
-\fr{2}{r K} r^4 \Big[r^8 -(b+3)r^4 -b\Big]\Big[r^8 -(11b+3)r^4
\nonumber \\
&& +3b(2b+5)\Big] \, \vp' -\fr{2}{(r^4-1)K} \Bigg[
m^2r^4\Big(r^4+b\Big)^2\Big[ r^{12} -(11b+14)r^8 +3\Big(2b^2 +4b
+3\Big)r^4  \nonumber \\
&& -b(10b+9) \Big] +8r^2\Big[ 6r^{16} +(15b-4)r^{12}
-b(11b+32)r^8 -b(2b+3)(4b-5)r^4  \nonumber \\
&& +b^2(2b+3) \Big] \Bigg] \, \vp \, = \, 0 .
 \label{SDEpsi}
 \eeqa
Here
 \beq
 K(r) = \Big(r^4+b\Big)\bigg[m^2r^2\Big(r^4+b\Big)^3
 -\Big(r^4-1\Big)\Big(7r^8 -14br^4 -b^2\Big)\bigg]
 \eeq
and the parameter $b$ is defined by
\begin{equation}
b = \sinh^2 \mu.
\end{equation}
In writing Eqs.~(\ref{SDEvphi}) and (\ref{SDEpsi}), we have used
that the background metric in Eq.~(\ref{pbrane}) is invariant up
to an overall multiplicative factor under rescalings of $r
\rightarrow r/\alpha$, $r_{\scriptscriptstyle{H}} \rightarrow
r_{\scriptscriptstyle{H}}/\alpha$, $t \rightarrow t/\alpha$,
$z^{i} \rightarrow z^{i}/\alpha$ for arbitrary constant $\alpha$.
Thus it should be understood that the dimensionless quantities in
equations above are actually $r \rightarrow
\bar{r}=r/r_{\scriptscriptstyle{H}}$ and $m \rightarrow
\bar{m}=r_{\scriptscriptstyle{H}} m$ with
$r_{\scriptscriptstyle{H}}= k^{1/(n-1)}$, respectively.

In order to see whether black D3-branes are unstable or not, now
we need to check if these two equations allow any spatially
regular solution outside the horizon for certain values of the
parameters $m$ and $b$. Let us first consider the boundary
behavior of the solutions near the horizon and at the infinity.
Since Eqs.~(\ref{SDEvphi}) and (\ref{SDEpsi}) are second-order
coupled linear differential equations for $\varphi$ and $\psi$,
there are four linearly independent mode solutions in general. The
asymptotic solutions at spatial infinity ( $i.e., r \sim \infty$)
are given by
 \beqa
 \varphi_{\infty} & \simeq & e^{\pm mr} u_{\pm} (r) \simeq e^{\pm mr}
r^{-5/2} \left( 1 + \frac{15}{8m} \frac{1}{r} + \cdots \right)
\nonumber \\
\psi_{\infty} & \simeq & e^{\pm mr} v_{\pm} (r) \simeq e^{\pm mr}
r^{-7/2} \left( \frac{1}{m} + \frac{35}{8m^2} \frac{1}{r} + \cdots
\right)
 \eeqa
up to overall arbitrary multiplicative constants. Note that only
half of these mode solutions are regular ones. Similarly, in the
vicinity of the event horizon ({\it i.e.}, $r= 1 + \triangle$), we
have two regular asymptotic mode solutions and two singular ones.
By being regular we mean that the solution should not produce any
curvature singularity at the horizon~\cite{Hirayama:2002hn}.
Asymptotic regular mode solutions are given by
 \beqa
\vp_{\scriptscriptstyle{I}\,r_{\scriptscriptstyle{H}}} & \sim & 1
-\fr{1}{L_1}
 \bigg[ (b+1)^5m^6 -16(10b-3)(b+1)^3m^4 -64(b+1)[b(109b-232)+75]m^2 \nonumber \\
 && \qquad\quad\,\, +8192b(2b-3) \bigg]
\Delta^2  +\cdots  \nonumber \\
\psi_{\scriptscriptstyle{I}\,r_{\scriptscriptstyle{H}}} & \sim & 1
+\fr{96}{L_1}(b+1)^2m^2 \left[ (b+1)^2m^2 +32(b-1)\right] \Delta
+\cdots
 \label{AsympsolHI}
 \eeqa
and
 \beqa
 \vp_{\scriptscriptstyle{II}\,r_{\scriptscriptstyle{H}}}
& \sim & \Delta +\fr{16}{L_1} \bigg[ (b+1)^4m^4
-2(27b-1)(b+1)^2m^2 -16[b(19b-82)+19] \bigg] \Delta^2 +\cdots \nonumber \\
\psi_{\scriptscriptstyle{II}\,r_{\scriptscriptstyle{H}}} & \sim &
-\fr{192}{L_1}(b+1)\left[(1 +b)^2m^2 +8(3b -1)\right] \Delta
-\fr{16}{L_1} \bigg[ (b+1)^4m^4 +2(9b-11)(b+1)^2m^2  \nonumber \\
&& +16[b(23b-122)+23] \bigg] \Delta^2 +\cdots .
 \label{AsympsolHII}
 \eeqa
Here
 \beq
 L_1 = 192(b+1)\left[(b+1)^2m^2 -8(3b -1)\right] .
 \eeq
Note that these asymptotic solutions become divergent as $L_1
\rightarrow 0$, {\it i.e.},
 \beq
 m^2 \simeq 8(3b-1)/(b+1)^2 .
 \label{DIdata}
 \eeq
To avoid this divergence for such values of parameters $m$ and $b$
(or, close to them), one may also use another set of asymptotic
solutions by linearly combining these solutions. They may be given
by
 \beqa
 \tilde{\vp}_{\scriptscriptstyle{I}\, r_{H}} & \sim & L_1
-\bigg[(b+1)^5m^6 -16(10b-3)(b+1)^3m^4 -64(b+1)[b(109b-232)+75]m^2  \nonumber \\
&& \qquad\quad\,\, +8192b(2b-3) \bigg] \Delta^2  +\cdots  \nonumber \\
\tilde{\psi}_{\scriptscriptstyle{I}\,r_{\scriptscriptstyle{H}}}
&\sim & L_1 +96(b+1)^2m^2 \left[ (b+1)^2m^2 +32(b-1)\right] \Delta
+\cdots
 \label{AsympsolHI2nd}
 \eeqa
and
 \beqa
 \tilde{\vp}_{\scriptscriptstyle{II}\,r_{\scriptscriptstyle{H}}} & \sim &
1 +\fr{1}{L_2} \bigg[ (b+1)^5m^6 -16(10b-3)(b+1)^3m^4
 -64(b+1)(b(109b-232)+75)m^2  \nonumber \\
&& \qquad\quad\,\, +8192b(2b-3) \bigg] \Delta +\cdots  \nonumber \\
\tilde{\psi}_{\scriptscriptstyle{II}\,r_{\scriptscriptstyle{H}}} &
\sim & 1 +\fr{1}{L_2} \bigg[ 7(b+1)^5m^6
+16(8b-21)(b+1)^3m^4 -64(b+1)(b(37b+24)-77)m^2 \nonumber \\
&& \qquad\quad\,\, +8192b(2b-3) \bigg] \Delta +\cdots  .
 \label{AsympsolHII2nd}
 \eeqa
Here
 \beq
 L_2 = 16 \left[ (b+1)^4m^4 -2(b+1)^2(27b-1)m^2 -16(b(19b-82)+19)
 \right].
 \eeq
The first pair is obtained simply by multiplying $L_1$ and the
second pair by linearly superposing two solutions such that the
term proportional to $\Delta^2$ in $\vp$ is regular, {\it i.e.},
being zero.

As in Ref.~\cite{Hirayama:2002hn}, let us consider two mode
solutions whose asymptotic behaviors near the horizon are those in
Eqs.~(\ref{AsympsolHI}) and (\ref{AsympsolHII}) as follows:
\begin{eqnarray}
 \label{solIvp}
 \varphi_{\scriptscriptstyle{I}} & \sim & \left\{
 \begin{array}{c}
   \vp_{\scriptscriptstyle{I}\,r_{H}} \\ \\
    A_{\scriptscriptstyle{I}} e^{-mr} u_{-}
    + B_{\scriptscriptstyle{I}}e^{mr} u_{+} ,
   \end{array}
\right.
 \quad \psi_{\scriptscriptstyle{I}} \sim \left\{
 \begin{array}{cl}
   \psi_{\scriptscriptstyle{I\,r_{\scriptscriptstyle{H}}}}  & \,\,\, {\rm for} \,\,
   r \rightarrow r_{\scriptscriptstyle{H}}  \\ \\
    \bar{A}_{\scriptscriptstyle{I}} e^{-mr} v_{-}
    + \bar{B}_{\scriptscriptstyle{I}}e^{mr} v_{+} & \,\,\, {\rm for} \,\,
    r \rightarrow \infty ,
   \end{array}
 \right.
\\
 \label{solIIvp}
 \varphi_{\scriptscriptstyle{II}} & \sim & \left\{
 \begin{array}{c}
    \vp_{\scriptscriptstyle{II\,r_{\scriptscriptstyle{H}}}}  \\ \\
    A_{\scriptscriptstyle{II}} e^{-mr} u_{-}
    +B_{\scriptscriptstyle{II}}e^{mr}u_{+} ,
    \end{array}
     \right.
 \psi_{\scriptscriptstyle{II}} \sim \left\{
 \begin{array}{cl}
    \psi_{\scriptscriptstyle{II\,r_{\scriptscriptstyle{H}}}} & {\rm for} \,\,
    r \rightarrow r_{\scriptscriptstyle{H}}  \\ \\
    \bar{A}_{\scriptscriptstyle{II}} e^{-mr} v_{-}
    +\bar{B}_{\scriptscriptstyle{II}}e^{mr}v_{+} & {\rm for} \,\, r \rightarrow
    \infty .
    \end{array}
\right.
\end{eqnarray}
Now any mode solutions that are regular at the horizon can be
written by
\begin{equation}
  \varphi = C\varphi_{\scriptscriptstyle{I}}
             +E\varphi_{\scriptscriptstyle{II}},
               \qquad
     \psi = C\psi_{\scriptscriptstyle{I}}
            +E\psi_{\scriptscriptstyle{II}}.
            \label{regsol}
\end{equation}
At the spatial infinity, they will behave like
\begin{eqnarray}
  &\varphi  \sim
  \left( CA_{\scriptscriptstyle{I}} +E A_{\scriptscriptstyle{II}} \right)
   e^{-mr} u_{-}(r) +\left( C B_{\scriptscriptstyle{I}} +E
   B_{\scriptscriptstyle{II}}\right) e^{mr} u_{+}(r),  \\
&\psi \sim  \left( C \bar{A}_{\scriptscriptstyle{I}} +E
   \bar{A}_{\scriptscriptstyle{II}} \right) e^{-mr}
   u_{-}(r) +\left( C \bar{B}_{\scriptscriptstyle{I}}
   +E\bar{B}_{\scriptscriptstyle{II}} \right)e^{mr}
   u_{+}(r) .
\end{eqnarray}
In order for these solutions to be regular, coefficients of the
exponentially growing parts should vanish
 \begin{equation}
C B_{\scriptscriptstyle{I}}+ E B_{\scriptscriptstyle{II}} =0,
 \qquad C \bar{B}_{\scriptscriptstyle{I}} + E
\bar{B}_{\scriptscriptstyle{II}} =0.
 \label{CriterionI}
\end{equation}
The condition that there exist any non-trivial coefficients $C$
and $E$ satisfying Eq.~(\ref{CriterionI}) will be
 \begin{equation}
 D(m,b) \equiv
 B_{\scriptscriptstyle{I}}\bar{B}_{\scriptscriptstyle{II}}
 -B_{\scriptscriptstyle{II}}\bar{B}_{\scriptscriptstyle{I}}
 =0
 \label{CriterionII}
\end{equation}
Namely, the existence of unstable mode solutions depends on
whether or not there are certain values of parameters $m$ and $b$
satisfying Eq.~(\ref{CriterionII}).

We have checked this numerically by using Mathematica. In more
detail, having given a certain form of initial data as in
Eqs.~(\ref{solIvp}) and (\ref{solIIvp}) near the horizon, we solve
the coupled equations Eqs.~(\ref{SDEvphi}) and (\ref{SDEpsi})
numerically, and evaluate the determinant $D(m,b)$ by using
numerical values for $\varphi_{\scriptscriptstyle I,II}$ and
$\psi_{\scriptscriptstyle I,II}$ at sufficiently large $r$. For a
given parameter $b$ of black D3-brane, we vary Kaluza-Klein mass
$m$ only and search for the $m^*$ at which $D(m^*,b) = 0$. This
can be achieved by finding out a $m^*$ around which $D(m,b)$
changes its sign. If there exists such $m^*$, the corresponding
solution is indeed the threshold unstable mode.

In the method described above, however, it should be pointed out
that in addition to real solutions there appear some fictitious
solutions in parameter space $(m,b)$ due to the divergence
behavior of initial data in Eqs.~(\ref{AsympsolHI}) and
(\ref{AsympsolHII}) for a certain range of parameters as in
Eq.~(\ref{DIdata}). One can find that these fictitious solutions
exactly coincide with the curve defined by Eq.~(\ref{DIdata}). By
using other set of initial data such as those in
Eqs.~(\ref{AsympsolHI2nd}) and (\ref{AsympsolHII2nd}), it turned
out that such fictitious solutions never appear.

\begin{figure}[tbp]
\resizebox{75mm}{5cm}{\includegraphics{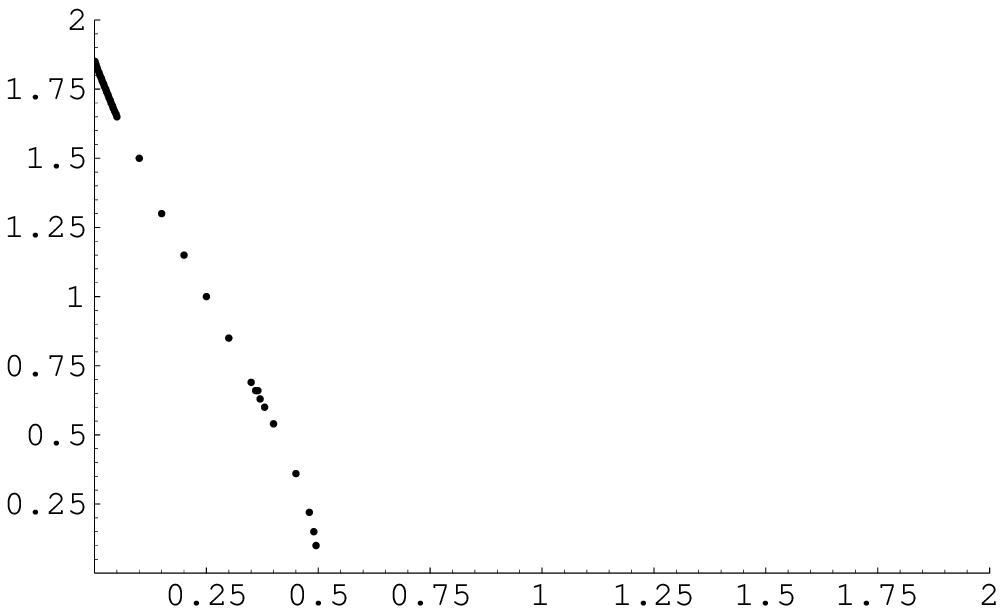}}
 \hfill
 \resizebox{75mm}{5cm}{\includegraphics{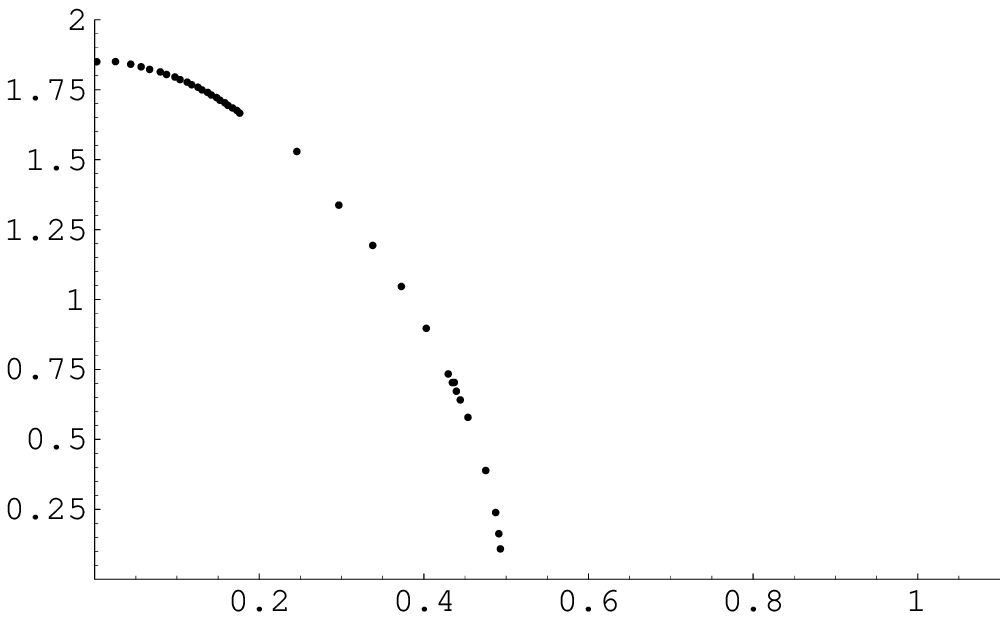}}
 \put(-247,-5){$b$}
 \put(-7,-5){$q$}
 \put(-455,150){$\bar{m}^*(b)$}
 \put(-205,150){$m^*(M,q)$}
 \caption{Behavior of dimensionless threshold masses for black
$D3$-branes at various values of $b$. Note that
$\bar{m}^*(b\!=\!0) \simeq 1.85$ for the uncharged case. On the
right hand side the threshold mass for the mass density $M=5$ is
plotted in terms of the extremality parameter $q$. The extremal
point is $q=1$ and the existence of solutions has been checked up
to $q \simeq 0.93$ ({\it i.e.}, $b \simeq 10$). Note that the
critical value at which instability disappears is $b_{\rm cr}
\simeq 0.50$ ({\it i.e.}, $q_{\rm cr} \simeq 0.49$). }
 \label{fig1}
\end{figure}

\begin{figure}[tbp]
\includegraphics{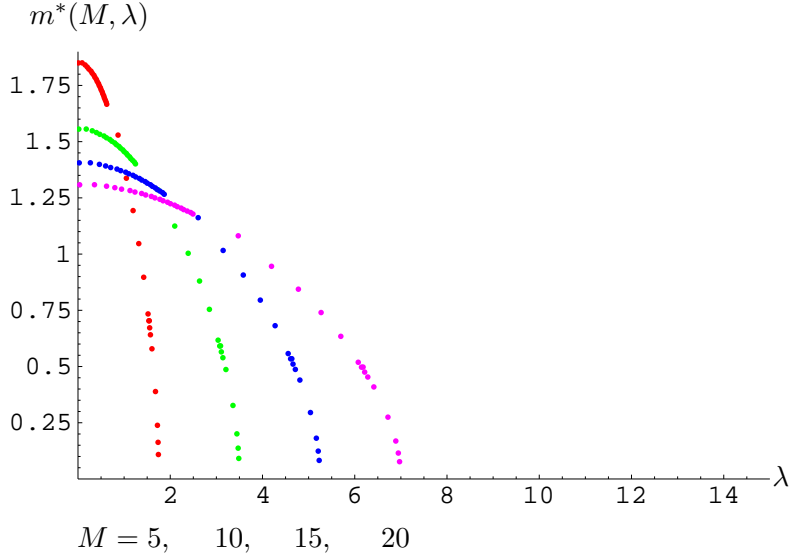}
 \put(-280,185){$m^*(M,\lambda)$}
 \put(1,10){$\lambda$}
 \put(-262,-13){$M=5,$}
 \put(-210,-13){$10,$}
 \put(-180,-13){$15,$}
 \put(-147,-13){$20$}
 \caption{Threshold masses vs. charge density $\lambda$ for black
D3-branes having various values of mass density $M$. Extremal
values of the charge density are $\lambda_{\rm max} =M/\sr{2}
\simeq 3.5,\, 7.1,\, 10.6,\, 14.1$, and critical values
corresponding to $b_{\rm cr} =1/2$ are $\lambda_{\rm cr}
=\sr{6}M/7 \simeq 1.8,\, 3.5,\, 5.3,\, 7.0$ for $M =5,\, 10,\,
15,\, 20$, respectively.}
 \label{fig2}
\end{figure}

Behaviors of some threshold masses we have obtained numerically in
parameter space are illustrated in Fig.~\ref{fig1} and
Fig.~\ref{fig2}. The diagram on the left hand side in
Fig.~\ref{fig1} shows how the dimensionless threshold mass
$\bar{m}^*(b)$ varies as the ``charge'' parameter $b$ increases.
Note that the uncharged black $D3$-brane is simply a product of
the seven-dimensional Schwarzschild black hole and
three-dimensional flat space and that our $s$-wave perturbation
analysis becomes equivalent to that of a pure gravitational
perturbation for such Schwarzschild black 3-brane since the gauge
perturbation $Q(r)$ becomes zero as the charge density $\lambda$
vanishes. In Fig.~\ref{fig1} the dimensionless threshold mass for
the uncharged black $D3$-brane turns out to be $\bar{m}^*(b=0)
\simeq 1.85$. We find such numerical value coincides with those
obtained in Refs.~\cite{Gregory:1994bj,Hirayama:2002hn} that, in
the uncharged limit, can be regarded as the result for the simple
Schwarzschild black 3-brane.

As the ``charge'' parameter $b$ increases, the dimensionless
threshold mass decreases monotonically, and approaches zero at $b
\simeq 0.50$. We have checked numerically that there is no
solution for larger values of $b$ than this critical value up to
about $b \simeq 10$ ({\it i.e.}, $q \simeq 0.93$), which is close
to the extremal point $q=1$. By using the scaling property
explained before we can obtain the actual threshold mass from our
results as
 \beq
 m^*(M,\lambda) = \bar{m}^*(b)/r_{\scriptscriptstyle{H}}
 = \left(\fr{5}{M}\right)^{1/4}m^*(r_{\scriptscriptstyle{H}}\!=\!1,q),
 \label{thresholdmass}
 \eeq
where $m^*(r_{\scriptscriptstyle{H}}\!=\!1,q) =
(5+4b)^{1/4}\bar{m}^*(b)/5^{1/4}$ and $b=b(q)$ should be
understood as a function of $M$ and $\lambda$ through
Eq.~(\ref{nonextrem}). The diagram on the right hand side in
Fig.~\ref{fig1} shows the behavior of the threshold mass for the
mass density $M=5$ fixed as the extremality parameter $q$
increases up to $q \simeq 0.93$. Therefore our numerical results
show that, although black $D3$-branes with a given mass density
are classically unstable under small $s$-wave perturbations when
they are charged weakly, as they get charged further this
instability decreases down to zero at a certain point far from the
extremal one. It can be also seen that the critical value of $q$
({\it i.e.}, $q_{\rm cr\,Num} \simeq 0.49$) at which the
instability disappears agrees very well with the predicted one
({\it i.e.}, $q_{\rm cr} = 2\sr{3}/7 \simeq 0.495$) in
Eq.~(\ref{criticalq}) through the GM conjecture.

Now let us consider the extremal black D3-brane, {\it i.e.}, $k
\rightarrow 0$ and $\mu$ (or, $b$) $\rightarrow \infty$ with $kb$
($\simeq k e^{2\mu}/4$) $=M/4 \equiv c$ fixed. The extremal black
D3-brane is expected to be stable since they correspond to the BPS
ground state in string theory. Our numerical result obtained up to
$q \simeq 0.93$ ({\it i.e.}, $b \simeq 10$) above also seems to
indicate that there would not appear instability mode when we
continue our analysis further up to the extremal point $q=1$.
However, $b \simeq 10$ is still small compared to $b \simeq
\infty$, and our analysis based on rescaled variables is not
appropriate to the case of extremal limit. In particular, the
initial data in Eqs.~(\ref{AsympsolHI}) and (\ref{AsympsolHII})
cannot be kept small as $b \rightarrow \infty$. Thus we study the
extremal case separately. By recovering the parameter $k$ and
taking the extremal limit in Eqs.~(\ref{SDEvphi}) and
(\ref{SDEpsi}), perturbation equations for the extremal $D3$-brane
are given by
\begin{eqnarray}
\mbox{} && \varphi'' +\frac{r^3}{P} \bigg[ 5c^4 m^2 + 5r^{14}
\Big(-7+m^2r^2\Big)
+ 5c^3 r^2\Big(3+4m^2r^2\Big) + c r^{10}\Big(37 + 20m^2r^2\Big) \nonumber \\
&& +3c^2r^6 \Big(21 +10m^2r^2\Big) \bigg]
\varphi' -\frac{m^2(c+r^4)^2}{P} \bigg[ c^3m^2 + r^{10}\Big(-7+m^2r^2\Big)
+3cr^6 \Big(4+m^2 r^2\Big) \nonumber \\
&& + c^2 r^2 \Big(11+3m^2 r^2\Big) \bigg] \varphi -\frac{2cr^5}{P}
\bigg[ 5c^2 +34cr^4 -7r^8 \bigg] \psi'  -\frac{2cr^2}{P} \bigg[
11c^3m^2-7m^2r^{12}  \nonumber \\
&& + c^2r^2 \Big(16 + 15m^2 r^2\Big) + cr^6\Big(80 -3m^2 r^2 \Big)
\bigg] \psi = 0
\end{eqnarray}
and
\begin{eqnarray}
 \mbox{} && \psi'' +\frac{r^3}{P} \bigg[ 5c^4m^2+ r^{14}\Big(-49 +5m^2r^2\Big)
 + c^3r^2\Big(-7+20m^2r^2\Big)+cr^{10}\Big(187+20m^2r^2\Big) \nonumber \\
&& +c^2r^6\Big(13+30m^2r^2\Big) \bigg] \psi' -\frac{1}{P} \bigg[
c^5 m^4 + m^2 r^{18}\Big(7+m^2r^2\Big)
+c^4m^2r^2\Big(21+5m^2r^2\Big)  \nonumber \\
&& + cr^{12} \Big(-384 -46m^2r^2 +5m^4r^4\Big) + 2c^2r^8\Big(16-46m^2r^2
+5m^4r^4\Big)  \nonumber \\
&& + 2c^3r^4\Big(16-9m^2r^2 + 5m^4r^4\Big) \bigg] \psi
-\frac{2r^5}{P} \bigg[ -6c^3 +17c^2r^4 -12cr^8 + r^{12} \bigg]
\varphi' \nonumber \\
&& -\frac{2m^2 r^2}{P} \Big(c+r^4 \Big)^2 \left(6c^2 -11cr^4
+r^8\right) \varphi  = 0 ,
\end{eqnarray}
where
\begin{eqnarray}
P(r) = r^4\Big(c+r^4\Big)\Big[ c^3m^2 + r^{10}\Big(-7+m^2r^2\Big)
+ cr^6\Big(14+3m^2r^2\Big) +c^2r^2\Big(1 + 3m^2r^2\Big) \Big] ,
\end{eqnarray}
and $m$ and $r$ are all dimensionful variables.

Differently from other extremal $p$-brane cases in
Ref.~\cite{Hirayama:2002hn}, these equations at the extremal point
are not decoupled. Repeating similar analysis as in the cases of
non-extremal black $D3$-branes, for given mass density $M$ (or,
$c$) we have checked for various values of $m$, but found no
regular solution. It confirms that the extremal $D3$-brane is
stable, at least under $s$-wave perturbations classically.

\section{Conclusions}

To conclude, we have investigated the classical stability of black
$D3$-branes in the type IIB supergravity under small
perturbations. For $s$-wave perturbations it turns out that black
$D3$-branes are unstable when they have small charge density. As
the charge density increases for a given mass density, however,
the instability decreases down to zero at a certain value of the
extremality parameter ({\it i.e.}, $q_{\rm cr} \simeq 0.49$), and
then black $D3$-branes become stable all the way down to the
extremal point. It has also been shown that such critical value at
which its stability behavior changes agrees very well with the
predicted one by the thermodynamic stability behavior of the same
system through the Gubser-Mitra conjecture. Therefore, although
the generalization of Reall's proof for the GM
conjecture~\cite{Reall:2001ag} to the case of black $D3$-branes
seems to be non-trivial as explained before, our direct comparison
confirms that the GM conjecture presumably holds to this case as
well.

The peculiar property in this theory that the five-form field
strength should be self-dual is imposed for linearized
perturbations as an additional constraint on the solution space.
Interestingly, however, it turned out that, for $s$-wave
perturbations, such constraint is automatically satisfied since
this constraint equation is precisely equal to one of linearized
equations for the gauge field. It should also be pointed out that
the $s$-wave instability for black $D3$-branes involves a
non-vanishing gauge field fluctuation. This property differs from
other cases of magnetically charged black $p$-branes studied in
Refs.~\cite{Gregory:1994bj,Hirayama:2002hn} where allowed $s$-wave
unstable solutions are for fluctuations of dilaton and
gravitational fields only with the gauge field frozen. Although
our stability analysis described in this paper is restricted to
$s$-wave perturbations so far, we expect that the essential
stability behavior of black $D3$-branes would not change even if
we consider non-$s$-wave perturbations further. It follows because
the strongest instability is expected to be carried by $s$-wave
perturbations~\cite{Gregory:1994bj}.\footnote{See also some other
argument in Ref.~\cite{Hirayama:2002hn} supporting such
expectation.}

\begin{figure}[tbp]
\includegraphics{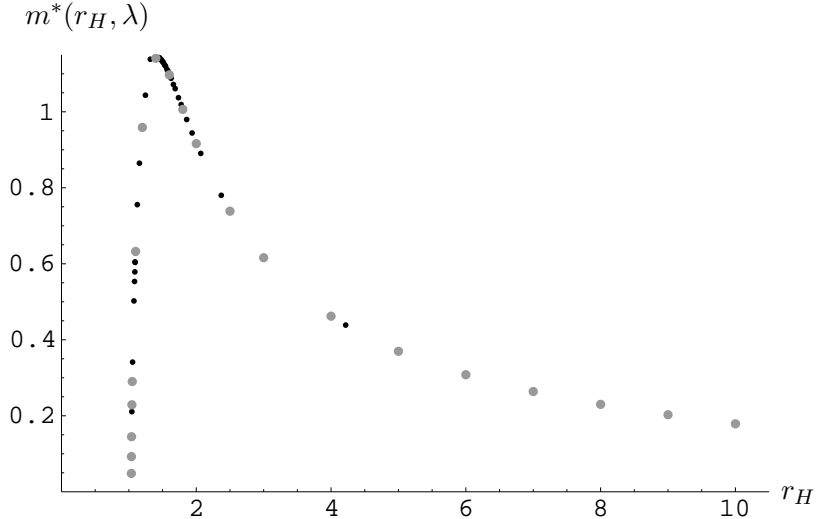}
\vspace{0.5cm}
 \put(-285,190){$m^*(r_H,\lambda)$}
 \put(1,10){$r_H$}
 \caption{Threshold masses with respect to the horizon radius for
 a given charge density $\lambda=2\sr{2}$. Black dots denote our
 result and gray dots that of Ref.~\cite{Gubser:2002yi}.
 $r_{\scriptscriptstyle{H}} =0$ corresponds to the extremal $D3$-brane,
 and the critical horizon radius is $r_{\scriptscriptstyle{H}{\rm cr}}
 =(2/\sr{3})^{1/4} \simeq 1.04$. }
 \label{fig3}
\end{figure}

Now let us compare our result with that of
Ref.~\cite{Gubser:2002yi} obtained by using the notion of
universality classes. Since the gauge field is frozen in the
dimensionally reduced action in Ref.~\cite{Gubser:2002yi}, it is
of interest to see how this difference in the stability analysis
affects to the result. In Fig.~\ref{fig3}, by using
Eq.~(\ref{thresholdmass}) and the relation between $b$ and the
horizon radius given by
 \beq
 r_{\scriptscriptstyle{H}} = \left(\lambda/\sr{2(n-1)b(b
 +1)}\right)^{1/(n-1)} ,
 \eeq
the threshold mass in our result is plotted with respect to the
horizon radius for a given charge density ({\it e.g.}, $\lambda
=2\sr{2}$) together with that in Ref.~\cite{Gubser:2002yi}. Here
$r_{\scriptscriptstyle{H}} =0$ corresponds to the extremal
$D3$-brane, and the critical horizon radius at which the
instability sets on ({\it i.e.}, $b_{\rm cr} =1/2$) is expected to
be $ r_{\scriptscriptstyle{H}{\rm cr}} =(2/\sr{3})^{1/4} \simeq
1.04$. Since the non-vanishing gauge fluctuation is proportional
to the charge density for small charge, we expect that the
assumption of frozen gauge fluctuation in
Ref.~\cite{Gubser:2002yi} might be fine when the charge density of
black $D3$-branes is small compared to the mass density. Thus
threshold masses in two methods probably agree very well at least
for large horizon radii as can be seen in Fig.~\ref{fig3}. Indeed
$r_{\scriptscriptstyle{H}} = 1.6$ corresponds to $q \simeq 0.1$
already. Interestingly, however, Fig.~\ref{fig3} shows that both
results are still in very good agreement even for horizon radii
near the critical value $r_{\scriptscriptstyle{H}} \simeq 1.04$
({\it i.e.}, $q \simeq 0.49$). Thus we see that the approximation
$\delta {\mathbf F}=0$ taken in Ref.~\cite{Gubser:2002yi} does not
change the results much even for rather ``strongly'' charged
branes.

\section*{Acknowledgments}

GK would like to thank T.~Hirayama, S.~Hyun, Y.~Lee, R.~M.~Wald
and P.~Yi for useful discussions. Authors also would like to thank
S.~S.~Gubser and A.~Ozakin for sending their numerical data for
threshold masses.

\newpage

\appendix

\section{Linearized perturbation equations}

For $s$-wave perturbations in the form of Eqs.~(\ref{PF}) and
(\ref{PH}) without imposing $\zeta =\eta =0$, the linearized
equations in Eqs.~(\ref{leomF}) and (\ref{leomG}) can be written
in components as follows:

\vspace{0.5cm}

\noindent
 ($tt$)-component:
 \beqa
 \mbox{} && U \varphi'' +\fr{U}{2} \left(
\ln r^{2n}U^3W^{-2} \right)' \varphi' +\left[ U''
+\fr{U'}{2}\left( \ln r^{2n}W^{-2}\right)'
-\fr{\lambda^2W^n}{2r^{2n}} -\fr{m^2}{W}
\right] \varphi  \nonumber  \\
&& -\fr{U'}{2} \psi' -\left[ U'' +\fr{U'}{2}\left( \ln
r^{2n}W^{-2}\right)' -\fr{\lambda^2W^n}{2r^{2n}} \right] \psi
+\fr{U'}{2} n\chi' +\fr{\lambda^2W^n}{2r^{2n}} n\chi \nonumber \\
&& +(n-2)\fr{U'}{2} \rho' +(n-2)\fr{\lambda^2W^n}{2r^{2n}} \rho
+\fr{U'}{2} \zeta' +\fr{\lambda^2W^n}{2r^{2n}} \zeta +\fr{m U'}{W
U} \eta -\fr{\lambda W^n}{r^{2n}} Q = 0
 \eeqa
($rr$)-component:
 \beqa
 \mbox{} && U \left[ \vp'' +n\chi'' +(n-2)\rho'' +\zeta'' \right]
 -\fr{U}{2} \left( \ln r^{2n}UW^{-2} \right)' \psi' - \left(
 \fr{\lambda^2W^n}{2r^{2n}} +\fr{m^2}{W} \right) \psi \nonumber \\
&& +\fr{3U'}{2} \vp' +\fr{\lambda^2W^n}{2r^{2n}} \vp +\fr{U}{2}
\left( \ln r^4UW^{-2} \right)' n\chi' +\fr{\lambda^2W^n}{2r^{2n}}
n\chi +(n-2)\fr{U}{2} \left( \ln UW^2 \right)'
\rho'  \nonumber \\
&& +(n-2)\fr{\lambda^2W^n}{2r^{2n}} \rho +\fr{U}{2} \left( \ln
UW^2 \right)' \zeta' +\fr{\lambda^2W^n}{2r^{2n}} \zeta +\fr{m}{W}
\left( 2\eta' -\fr{U'}{U}\eta \right) -\fr{\lambda W^n}{r^{2n}} Q
= 0
 \eeqa
($\theta\theta$)-component:
 \beqa
 \mbox{} && U\chi'' +\fr{U}{2} \left(\ln r^{4n}U^2W^{-n-2}\right)'
 \chi' +\Bigg[ U\fr{\left(r^2W^{-1}\right)''}{r^2W^{-1}} +\fr{U}{2}\left(\ln
 r^2W^{-1}\right)' \left(\ln r^{2(n-2)}U^2\right)' \nonumber \\
 && -(n-2)\fr{\lambda^2W^n}{2r^{2n}}
 -\fr{m^2}{W} \Bigg] \chi +\fr{U}{2}\left(\ln r^2W^{-1}\right)' \vp'
 -\fr{\lambda^2W^n}{2r^{2n}}\vp -\fr{U}{2}\left(\ln
 r^2W^{-1}\right)' \psi' \nonumber \\
&& -\Bigg[ U\fr{\left(r^2W^{-1}\right)''}{r^2W^{-1}}
+\fr{U}{2}\left(\ln
 r^2W^{-1}\right)' \left(\ln r^{2(n-2)}U^2\right)'
 +\fr{\lambda^2W^n}{2r^{2n}} \Bigg] \psi +(n-2)\fr{U}{2}\left(\ln r^2W^{-1}\right)'
\rho' \nonumber \\
&& -(n-2)\fr{\lambda^2W^n}{2r^{2n}} \rho +\fr{U}{2}\left(\ln
r^2W^{-1}\right)'
 \zeta' -\fr{\lambda^2W^n}{2r^{2n}} \zeta +\fr{m}{W}\left(\ln
r^2W^{-1}\right)' \eta +\fr{\lambda W^n}{r^{2n}} Q = 0
 \eeqa
($z^i=z^j$)-component:
 \beqa
 \mbox{} && U\rho'' +\fr{U}{2}\left(\ln r^{2n}U^2W^{n-4}\right)'
 \rho' +\Bigg[ U\fr{W''}{W} +\fr{U}{2}\fr{W'}{W}\left(\ln
 r^{2n}U^2W^{-4}\right)' +(n-4)\fr{\lambda^2W^n}{2r^{2n}} -\fr{m^2}{W} \Bigg]
 \rho  \nonumber \\
&& +\fr{U}{2}\fr{W'}{W} \vp' +\fr{\lambda^2W^n}{2r^{2n}} \vp
-\fr{U}{2}\fr{W'}{W} \psi' -\left[ U\fr{W''}{W}
+\fr{U}{2}\fr{W'}{W}\left(\ln r^{2n}U^2W^{-4}\right)'
-\fr{\lambda^2W^n}{2r^{2n}} \right] \psi \nonumber \\
&& +\fr{U}{2}\fr{W'}{W} n\chi' +\fr{\lambda^2W^n}{2r^{2n}} n\chi
+\fr{U}{2}\fr{W'}{W} \zeta' +\fr{\lambda^2W^n}{2r^{2n}} \zeta
+\fr{m}{W}\fr{W'}{W} \eta -\fr{\lambda W^n}{r^{2n}} Q = 0
 \eeqa
($rz^i$)-component:
 \beqa
 \mbox{} && U\left[\vp' +n\chi' +(n-3)\rho'\right] +\fr{U}{2}\left(\ln UW^{-1}\right)' \vp
 -\fr{U}{2}\left(\ln r^{2n}UW^{-3}\right)'\psi  \nonumber \\
&& +\fr{U}{2}\left(\ln r^2W^{-2}\right)'n\chi +\fr{1}{m} \left[
U\fr{W''}{W} +\fr{U}{2}\fr{W'}{W}\left(\ln r^{2n}U^2W^{-4}\right)'
-\fr{\lambda^2W^n}{r^{2n}} \right] \eta = 0
 \label{leomri}
 \eeqa
($z^i\not= z^j$)-component:
 \beqa
  \mbox{} && U\zeta'' +\fr{U}{2}\left(\ln r^{2n}U^2W^{-2}\right)'
  \zeta' +\left[ U\fr{W''}{W} +\fr{U}{2}\fr{W'}{W}\left(\ln r^{2n}U^2W^{-4}\right)'
-\fr{\lambda^2W^n}{r^{2n}} \right] \zeta  \nonumber \\
&& +\fr{m}{W} \left[ 2\eta' +\left(\ln r^{2n}W^{-4}\right)' \eta
\right] -\fr{m^2}{W}\left[\vp +\psi +n\chi +(n-4)\rho\right] = 0
 \label{leomiNj}
 \eeqa
$\delta F$:
 \beqa
 \lambda \left[ \vp +\psi +(n-2)\rho -n\chi +\zeta \right] -2Q &=& 0
 \label{leomFcomp1} \\
 \lambda \left[ \vp' +\psi' +(n-2)\rho' -n\chi' +\zeta' \right] -2Q' &=& 0
 \label{leomFcomp2}
 \eeqa

Two non-trivial components of the equation $\nabla_M h^{MN} =
\fr{1}{2} \nabla^N h$ can be written as

\vspace{0.3cm}

\noindent
 ($r$)-component:
 \beqa
 \mbox{} &&
 U\left[\vp' -\psi' +n\chi' +(n-2)\rho' +\zeta'\right] +U' \vp
 -U\left(\ln r^{2n}UW^{-2}\right)' \psi  \nonumber \\
&& +U\left(\ln r^{2}W^{-1}\right)'n\chi +U\fr{W'}{W} \left[
(n-2)\rho +\zeta \right] +\fr{2m}{W}\eta = 0
 \label{TGr}
 \eeqa
($z^i$)-component:
 \beq
 m\left[\vp +\psi +n\chi +(n-4)\rho -\zeta\right] -2\eta' -\left(\ln
 r^{2n}W^{-2}\right)'\eta = 0
 \label{TGi}
 \eeq
Note that Eqs.~(\ref{leomri}) and (\ref{leomiNj}) do not contain
the function $Q(r)$, and also that they are related to
Eqs.~(\ref{TGr}) and (\ref{TGi}) as
 \beqa
 m(\tilde{\ref{leomri}}) &=& \fr{m}{2}(\tilde{\ref{TGr}}) +\fr{U}{2}(\tilde{\ref{TGi}})'
 -\fr{mU}{2}\fr{W'}{W} \left[\vp -\psi +n\chi +(n-2)\rho\right] \nonumber \\
&& +U\eta'' +\fr{U}{2}\left(\ln r^{2n}W^{-2}\right)' \eta' +\Bigg[
\fr{U}{2}\fr{W'}{W}\left(\ln r^{2n}U^2W^{-2}\right)' -n\fr{U}{r^2}
-\fr{\lambda^2W^n}{r^{2n}}  \nonumber \\
&& -\fr{m^2}{W} \Bigg] \eta -\fr{mU}{2}\fr{W'}{W} \zeta
 \label{leomTG} \\
(\tilde{\ref{leomiNj}}) &=& -\fr{m}{W} (\tilde{\ref{TGi}})
+U\zeta'' +\fr{U}{2}\left(\ln r^{2n}U^2W^{-2}\right)' \zeta' +\Bigg[ U\fr{W''}{W} \nonumber \\
&& +\fr{U}{2}\fr{W'}{W}\left(\ln r^{2n}U^2W^{-4}\right)'
-\fr{\lambda^2W^n}{r^{2n}} -\fr{m^2}{W} \Bigg] \zeta
-\fr{2m}{W}\fr{W'}{W} \eta ,
 \eeqa
where $(\tilde{A}l)$ with $l=5,6,8,9$ denotes the left hand side
of Eq.~($Al$). Thus one finds that in the Reall gauge ({\it i.e.},
$\zeta =\eta =0$) the linearized perturbation equation
Eq.~(\ref{leomiNj}) becomes equivalent to the $z^i$-component of
the transverse gauge Eq.~(\ref{TGi}). As can be seen in
Eq.~(\ref{leomTG}), however, Eq.~(\ref{leomri}) differs from the
$r$-component of the transverse gauge Eq.~(\ref{TGr}) as the black
brane gets charged ({\it i.e.}, $W' \not= 0$). This property is
different from the cases studied in
Refs.~\cite{Hirayama:2002hn,Reall:2001ag} where some of linearized
perturbation equations in the Reall gauge become exactly equal to
the transverse gauge.

\section{Perturbation equations for the $D3$-brane}

For black $D3$-branes ({\it i.e.}, $n=D/2=p+2=5$), the linearized
perturbation equations above in the Reall gauge $\zeta =\eta =0$
become

\vspace{0.5cm}

\noindent
 ($tt$)-component:
 \beqa
 \mbox{} && \left(1 -\fr{1}{r^4}\right) \varphi'' +\fr{5r^4 +6b+1}{r^5(1+b/r^4)} \varphi'
 -\fr{m^2(r^{12} +3br^8 +3b^2r^4 +b^3) -4b(b+1)r^2}{r^{12}(1 +b/r^4)^{2}} \varphi \nonumber  \\
&& -\fr{(b+2)r^4 +b}{r^9(1 +b/r^4)} \psi' -\fr{4b(b+1)}{r^{10}(1 +
b/r^4)^{2}} \psi +\fr{(b+2)r^4 +b}{r^9(1 + b/r^4)} 5\chi' +\fr{4b(b+1)}{r^{10}(1
+ b/r^4)^{2}} 5\chi  \nonumber \\
&& +\fr{(b+2)r^4 +b}{r^9(1 + b/r^4)} 3\rho'
+\fr{4b(b+1)}{r^{10}(1+ b/r^4)^{2}} 3\rho
-\fr{2\sr{2b(b+1)}}{r^{10}(1 + b/r^4)^{2}} Q = 0
 \eeqa
($rr$)-component:
 \beqa
 \mbox{} && \left(1 -\fr{1}{r^4}\right) \left[ \vp'' +5\chi'' +3\rho'' \right]
 -\fr{5r^8 +(4b-3)r^4 -2b}{r^9(b+1/r^4)} \psi' -\fr{m^2(r^4+b)^3 +4b(b+1)r^2}{r^{12}(1
 + b/r^4)^{2}} \psi \nonumber \\
&& +3\fr{(b+2)r^4 +b}{r^9(1 + b/r^4)} \vp' +\fr{4b(b+1)}{r^{10}(1
+ b/r^4)^{2}} \vp
+\fr{2r^8 +br^4 +b}{r^9(1 + b/r^4)} 5\chi' +\fr{4b(b+1)}{r^{10}(1 +b/r^4)^{2}} 5\chi \nonumber \\
&& +\fr{(3b+2)r^4 -b}{r^9(1 + b/r^4)} 3\rho'
+\fr{4b(b+1)}{r^{10}(1 + b/r^4)^{2}} 3\rho
-\fr{2\sr{2b(b+1)}}{r^{10}(1 + b/r^4)^{2}} Q = 0
 \eeqa
($\theta\theta$)-component:
 \beqa
 \mbox{} && \left(1 -\fr{1}{r^4}\right) \chi'' +\fr{10r^8
 +(5b-6)r^4 -b}{r^9(1+b/r^4)} \chi' -\fr{1}{r^{12}(1+b/r^4)^{2}}\Bigg[ m^2(r^{12}
 +3br^8 +3b^2r^4 +b^3) \nonumber \\
 && -8r^{10} -16br^6 +4b(3b+5)r^2 \Bigg] \chi +\fr{1 -1/r^4}{r(1 +b/r^4)} \vp'
 -\fr{4b(b+1)}{r^{10}(1 + b/r^4)^{2}} \vp -\fr{1 -1/r^4}{r(1 +b/r^4)} \psi' \nonumber \\
&& -4\fr{2r^8 +4br^4 +b(b-1)}{r^{10}(1 + b/r^4)^{2}} \psi +\fr{1
-1/r^4}{r(1 +b/r^4)} 3\rho' -\fr{4b(b+1)}{r^{10}(1 + b/r^4)^{2}} 3\rho \nonumber \\
&& +\fr{2\sr{2b(b+1)}}{r^{10}(1 + b/r^4)^{2}} Q = 0
 \eeqa
($z^i=z^j$)-component:
 \beqa
 \mbox{} && \left(1 -\fr{1}{r^4}\right) \rho'' +\fr{5r^8 +(8b-1)r^4 -4b}{r^9(1 +b/r^4)} \rho'
 -\fr{m^2(r^{12} +3br^8 +3b^2r^4 +b^3) -12b(b+1)r^2}{r^{12}(1 + b/r^4)^{2}} \rho  \nonumber \\
&& +b \fr{1 -1/r^4}{r^5(1 +b/r^4)} \vp' +\fr{4b(b+1)}{r^{10}(1
+b/r^4)^{2}} \vp -b \fr{1 - 1/r^4}{r^5(1 +b/r^4)} \psi'
-\fr{4b(b+1)}{r^{10}(1 + b/r^4)^{2}} \psi \nonumber \\
&& +b \fr{1 - 1/r^4}{r^5(1 +b/r^4)} 5\chi' +\fr{4b(b+1)}{r^{10}(1+
b/r^4)^{2}} 5\chi -\fr{2\sr{2b(b+1)}}{r^{10}(1 + b/r^4)^{2}} Q =0
 \eeqa
($rz^i$)-component:
 \beqa
 \mbox{} && \vp' +5\chi' +2\rho' +\fr{2}{r^5(1-1/r^4)} \vp \nonumber \\
&&  -\fr{5r^8 +3(b-1)r^4 -b}{r^9(1-1/r^4)(1+b/r^4)} \psi +\fr{r^4
-b}{r^5(1+b/r^4)} 5\chi = 0
 \label{leomriRG}
 \eeqa
($z^i\not= z^j$)-component:
 \beqa
 \vp +\psi +5\chi +\rho = 0
 \label{leomiNjRG}
 \eeqa
$\delta F$:
 \beq
 \lambda \left[ \vp +\psi +3\rho -5\chi \right] -2Q = 0
 \label{leomFcompRG}
 \eeq
Here $b= \sinh^2\mu$ and we set $k=1$.

\end{document}